# Recupero di un raro banco ottico del Melloni costruito nella Palermo della "belle époque"

**Aurelio Agliolo Gallitto[1], Salvatore Licata[2], Filippo Mirabello[1], Francesca Taormina[3]**

[1]Dipartimento di Fisica e Chimica, Università di Palermo, via Archirafi 36, 90123 Palermo

[2]Liceo Classico Statale "Umberto I", via Filippo Parlatore 26/c, 90145 Palermo

[3]Osservatorio Astronomico "G.S. Vaiana", Piazza Parlamento 1, 90100 Palermo

# Recovery of a rare Melloni's optical bench built in Palermo during the "belle époque"

**Abstract.** In the article, we will report on the recovery of a Melloni's optical bench built at the end of 1800 by the "macchinista" Filippo Caliri in the "belle époque" of Palermo. A scientific instrument of particular historical and didactic interest belonging to the collection of Liceo Classico Statale "Umberto I" of Palermo. In the article, we will discuss the technical aspects of the interventions carried out.

*Keywords*: scientific and technological heritage, historical physics instruments

**Riassunto.** In questo articolo discuteremo del recupero di uno strumento scientifico di particolare interesse storico-didattico appartenente alla Collezione del Liceo Classico Statale "Umberto I" di Palermo: un raro banco ottico del Melloni costruito alla fine del 1800 nella Palermo della "belle époque" dal "macchinista" Filippo Caliri. Nell'articolo discuteremo gli aspetti tecnici degli interventi conservativi effettuati.

*Parole chiave*: patrimonio scientifico e tecnologico, strumenti storici di fisica

**Corresponding Author:**

Aurelio Agliolo Gallitto
Dipartimento di Fisica e Chimica
via Archirafi 36, 90123 Palermo
Tel: 091.23891702, Fax: 091.6162461
E-mail: aurelio.agliologallitto@unipa.it

# 1. Introduzione

Il Liceo Classico Statale "Umberto I" di Palermo fu istituito con Regio Decreto del 20 giugno 1878 (Lo Scrudato, 2015), ma le sue origini risalgono al 1788 nel Convento di Sant'Anna, dove avevano sede le cosiddette Scuole Normali gestite da enti ecclesiastici, un'innovazione rispetto all'istruzione tradizionalmente riservata agli aristocratici. La Scuola Normale di Sant'Anna, nel novembre 1860, diventò il Ginnasio di Sant'Anna per volontà del generale Giuseppe Garibaldi. Dal 1878 il nome venne mutato in Ginnasio "Principe Umberto" e successivamente in Regio Liceo Ginnasiale "Umberto I". L'Istituto fu positivamente investito dalla rivoluzione culturale e scientifica della "belle époque" palermitana. Illustri fisici e matematici vi insegnarono, quali Michele Cantone (1857-1932) che successivamente divenne direttore dell'Istituto di Fisica dell'Università di Napoli e Giovanni Maisano (1851-1929) che successivamente ottenne la cattedra di Algebra Complementare all'Università di Palermo. Tra il 1924 e il 1927 parte di significativi finanziamenti inviati dal Ministero furono destinati all'acquisto di materiale scientifico per i gabinetti di Fisica e di Chimica.

Oggi il Liceo possiede una ricca collezione di strumenti scientifici di interesse storico che include strumenti di meccanica, acustica, termologia, elettrostatica, elettromagnetismo e ottica. A quest'ultima categoria appartiene il banco ottico del Melloni, descritto nel presente articolo. Il banco ottico è stato rinvenuto per caso in un magazzino di "rottami" assieme a una "Wommelsdorfsche Kondensatormaschine" e stava appunto per essere discaricato assieme ad altro materiale in disuso.

In questo articolo, discuteremo gli aspetti tecnici degli interventi effettuati sul banco ottico della Collezione del Liceo "Umberto I". Le attività sono state condotte nell'ambito di uno Stage Formativo organizzato dal Dipartimento di Fisica e Chimica dell'Università di Palermo presso l'Officina di Restauro di Strumenti Antichi. In tale occasione, due studenti del Liceo, sotto la guida di esperti e docenti universitari, hanno collaborato alle attività laboratoriali, maturando un primo approccio di natura didattica alle problematiche degli strumenti scientifici di interesse storico-didattico.

# 2. Il banco ottico del Melloni

Il banco ottico del Melloni fu messo a punto nella prima metà del XIX secolo da Macedonio Melloni (1798-1854), per condurre esperienze riguardanti il calore raggiante, termine con il quale veniva solitamente indicata la radiazione infrarossa. I manuali di Fisica dell'Ottocento riportano vari esperimenti che possono essere realizzati con il banco ottico, come l'esperimento sulla propagazione rettilinea del calore o la verifica della legge dell'inverso del quadrato delle distanze, per dimostrare l'analogia tra luce e calore (Ganot, 1863).

Il banco ottico del Melloni è composto da una guida di ottone, sulla quale è incisa una scala graduata, fissata su una base di legno montata su quattro viti calanti. La guida funge da sostegno sia per la sorgente di calore (radiazione infrarossa) sia per il rivelatore. Nel banco ottico venivano solitamente usate le seguenti sorgenti di calore.

- Lampada a olio di Locatelli, con sistema di alimentazione dello stoppino a flusso costante.
- Cubo di Leslie, costituito da una scatola cubica di ottone contenente acqua portata a ebollizione da un bruciatore ad alcol alloggiato nella parte sottostante; il cubo ha una faccia annerita, una bianca, una lucidata a specchio e una smerigliata per variare l'emittanza media delle superfici, massima per il nero e minima per il bianco.
- Spirale di platino portata a incandescenza con un bruciatore ad alcol e negli apparati più recenti elettricamente.
- Lamine di rame riscaldate a temperatura elevata, per esempio con un bruciatore ad alcol.

Il banco ottico è corredato da vari accessori, composti da lastrine di vetri colorati, schermi per selezionare il fascio generato dalle varie sorgenti e fenditure per regolarne l'intensità. Per esempio, un doppio diaframma (chopper) costituito da due pannelli neri paralleli distanziati di circa un centimetro posto su una base reclinabile che ha funzione di "interruttore", poiché alzato interrompe il fascio assorbendo la radiazione; oppure, un diaframma con fenditure di varie forme e dimensioni che possono essere aperte una per volta per mezzo di un meccanismo a disco rotante. Infine, l'elemento fondamentale del banco ottico del Melloni è il rivelatore, costituito da una termopila e un galvanometro. La termopila, formata da lamine di bismuto e antimonio opportunamente collegate in serie, fu ideata nel 1829 dal fisico italiano Leopoldo Nobili (1784-1835) e si basa sull'effetto termoelettrico scoperto nel 1822 dal fisico estone Thomas Johann Seebeck (1770-1831). In fig. 1 è riportata una illustrazione del banco ottico del Melloni (Ganot, 1863), in cui sono mostrati i vari elementi che lo compongono.

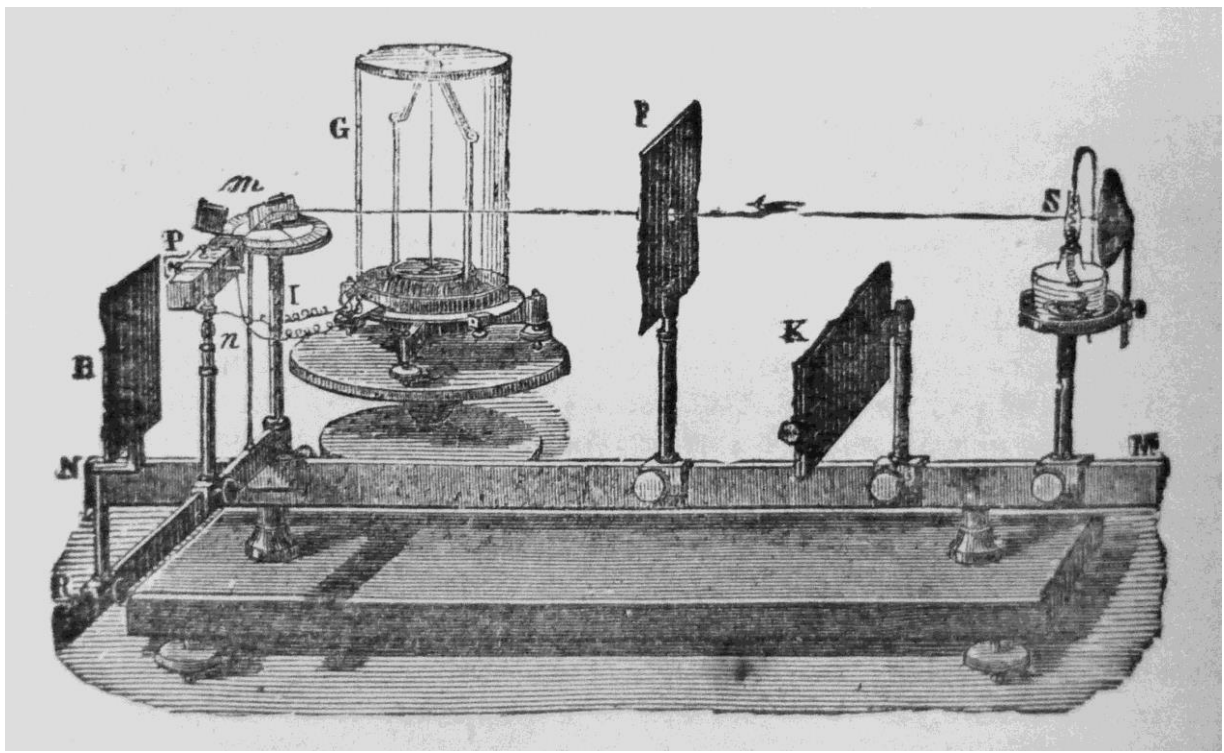

**Fig. 1.** Illustrazione del banco ottico del Melloni, tratta da Ganot. Sono mostrati i seguenti elementi: S – sorgente, K – diaframma, F – diaframma, I – goniometro con specchio (m), R – alidata, P – pila termo-elettrica (termopila), G – galvanometro, H – diaframma a gomito.

Lo strumento oggetto del restauro è costituito da una base rettangolare in legno di mogano, 20 cm larga e 70 cm lunga, poggiante su quattro piedi in ottone laccato provvisti di viti calanti. Esso è identificato da un numero di inventario principale 8641 riportato su un'etichetta cartacea, congiuntamente alle due frasi "Apparato Melloni – per calore raggiante". Inoltre, sulla base di legno è presente una targhetta metallica con l'incisione della firma del costruttore "Filippo Caliri – Palermo" (fig. 2). Sulla base di legno è collocata la guida di ottone laccato, con una scala incisa che misura 102 cm, su cui possono essere posizionati i cavalieri in ottone laccato muniti di viti a pressione. Sono pervenuti in totale sei cavalieri, tre dei quali sono numerati secondo la sequenza: IV 90, IV 92 e IV 93. Considerato che sui restanti cavalieri si trovano tracce della stessa vernice, è probabile che essi riportassero analoga numerazione, oggi non più leggibile. Anche la base di legno riporta il numero IV 88 applicato con la stessa vernice. I cavalieri reggono i seguenti elementi.

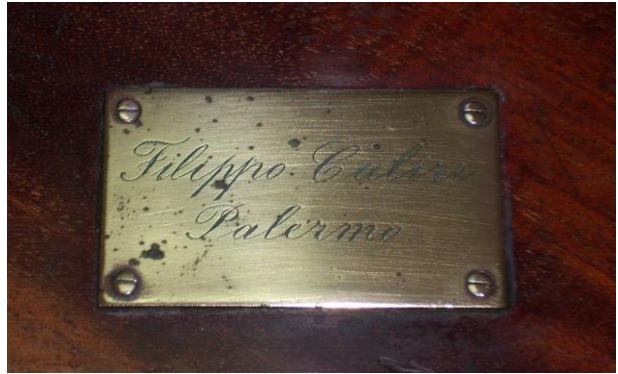

**Fig. 2.** Targhetta con inciso "Filippo Caliri Palermo".

- Uno regge quattro lampade a incandescenza con portalampada in vetro (fig. 3) collocate sopra un supporto orizzontale disposto perpendicolarmente alla guida del banco ottico (fig. 4); uno regge una sola lampada con portalampada in vetro; le due disposizioni di lampade probabilmente servivano ad avere sorgenti di differente intensità luminosa.
- Tre cavalieri reggono altrettanti schermi di forma rettangolare, uno dei quali dotato di un diaframma con finestre circolari, utilizzato per selezionare ed eventualmente variare l'intensità del fascio generato dalla sorgente.
- Infine, sul sesto cavaliere è fissata un'alidata che sostiene un goniometro graduato con puntatore, sul quale veniva posizionato uno specchio per deviare il fascio radiante di uno specifico angolo.

Due degli schermi riportano una etichetta cartacea, sulla quale è riportato il numero di inventario principale, 8641. La presenza di diversi numeri di inventario, lascia intuire che gli accessori siano stati aggiunti in periodi differenti e che lo strumento constava originariamente di più parti, compresa la termopila e il galvanometro per la rivelazione della radiazione infrarossa.

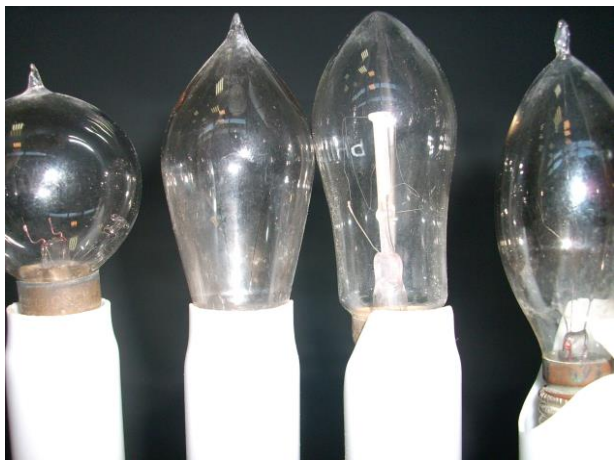

**Fig. 3.** Lampade a incandescenza usate come sorgente di radiazione infrarossa.

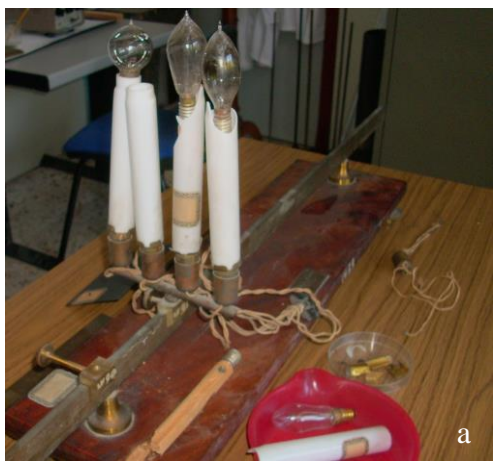

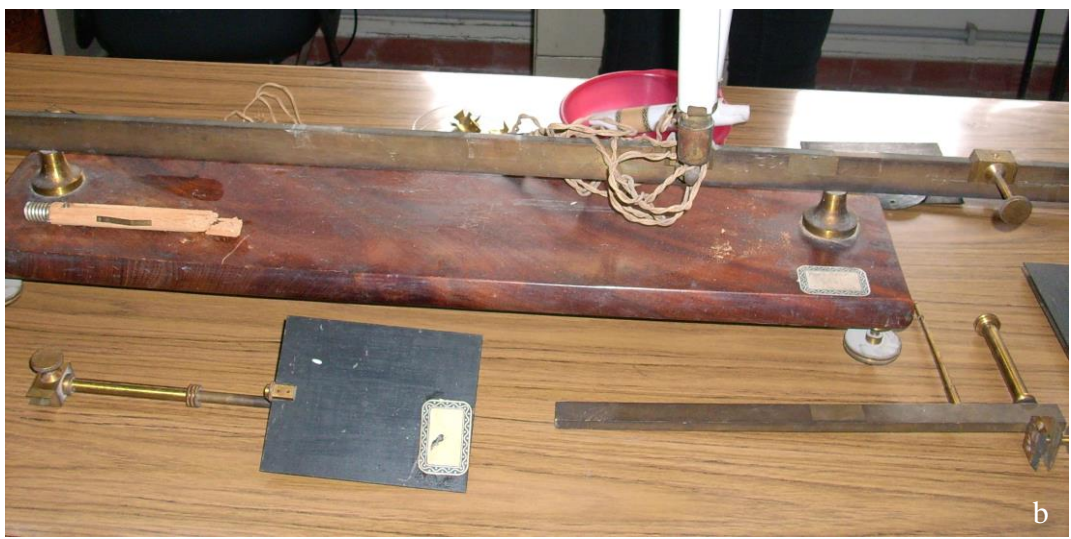

**Fig. 4.** Foto di alcuni particolari del banco ottico prima degli interventi di restauro.

## 3. Contesto storico: la figura di Filippo Caliri

L'interesse di approfondire il contesto storico nasce principalmente dalla volontà di indagare sulla figura di Filippo Caliri (1837-1922), nome del costruttore, che compare sulla targa metallica applicata sulla base di legno del banco ottico (fig. 2).

Laureatosi in medicina nel 1859, Caliri fu "chirurgo militare di battaglione" durante le campagne del periodo garibaldino, che portarono all'Unità d'Italia. Conclusa l'esperienza militare, decide di intraprendere una carriera nell'ambito della Fisica sperimentale e applicata. Infatti, dall'elenco dei docenti della Regia Università di Palermo (1820-1880) e dall'Annuario della Istruzione Pubblica del 1861/62, si trova che Filippo Caliri ricoprì il ruolo di dimostratore della cattedra di Fisica Sperimentale nel 1861/62 (Direttore del Gabinetto di Fisica Prof. Giuseppe Lo Cicero), fu assistente nel 1864 e macchinista nel 1870/71 (Direttore del Gabinetto di Fisica Prof. Pietro Blaserna ed assistente provvisorio Damiano Macaluso). Con Macaluso, Caliri pubblicò nel 1871 un articolo sul *Giornale di Scienze Naturali ed Economiche* dal titolo "Misure barometriche delle altezze sul livello del mare di talune montagne che circondano Palermo" (Caliri & Macaluso, 1871). Il valore scientifico di Caliri è confermato dalla nomina come supplente nell'insegnamento universitario dall'illustre Pietro Blaserna e come ordinario di Fisica nei RR Istituti Tecnici. Viene commemorato e ricordato da Merenda come uno scienziato che « *alle ricerche dottrinali preferiva le applicazioni* ». Questa sua passione per la Fisica applicata lo condusse alla realizzazione di un nuovo telegrafo elettrico stampante, in sostituzione di quello di Morse a linee e punti, invenzione che tuttavia non godrà di grande successo. Caliri è ricordato soprattutto per la fondazione nel 1870 dell'Officina Tecnica Piazzi, dove lavorerà per 22 anni come scienziato e meccanico realizzando strumenti di Fisica per i Gabinetti delle Scuole e delle Università (Merenda, 1924). Si dimostrò sensibile al fervore artistico e culturale del periodo. Viene infatti menzionato tra i professionisti e studiosi, attivamente coinvolti nelle conferenze organizzate dal Casino delle Arti di Palermo, fondato nel 1864, su temi attinenti le arti e l'industria (Sessa, 2008). Filippo Caliri andò in pensione con decreto del 8/5/1913 (Gazzetta Ufficiale, 1914).

Questa breve digressione storica sulla figura di Caliri consente di datare approssimativamente il banco ottico tra la fine del XIX e gli inizi del XX secolo e di confermare la sua produzione locale: ciò conferisce un valore di rarità allo strumento e valorizza maggiormente la collezione a cui esso appartiene.

## 4. Interventi conservativi

Gli interventi a cui è stato sottoposto il banco ottico della collezione del Liceo "Umberto I" hanno riguardato diverse problematiche di conservazione, legate principalmente alla natura polimaterica dello strumento. Considerato il pregio di tale apparato, abbiamo effettuato interventi conservativi mirati a preservare tutti gli elementi restanti e recuperare l'aspetto originario, in modo da facilitare la comprensione del principio di funzionamento da parte del pubblico meno esperto e contribuire in questo modo alla valorizzazione dello strumento (Marotti, 2004; Miniati, 1988). Sebbene alcuni elementi fossero mancanti, lo strumento nel complesso versava in buono stato di conservazione e quindi sono stati effettuati i seguenti interventi.

- Rimozione del deposito superficiale con l'uso di pennellesse.
- Pulitura superficiale della base in legno con soluzione del tensioattivo non ionico a pH neutro Tween 20 in acqua demineralizzata a una concentrazione del 2%.
- Pulitura degli elementi in ottone laccato e dei tre schermi metallici con petrolio bianco e panno di cotone.
- Ricostruzione dei supporti in legno interni ai portalampada e pulitura delle parti in vetro con alcol.
- Ultimato il processo di pulitura, la base in legno è stata protetta con olio specifico per il restauro del legno.

Le foto dell'apparato prima e dopo gli interventi di restauro sono riportate rispettivamente in fig. 4 e in fig. 5. Conclusi gli interventi, lo strumento è stato collocato nelle sale espositive del Liceo per essere fruito durante le visite della collezione.

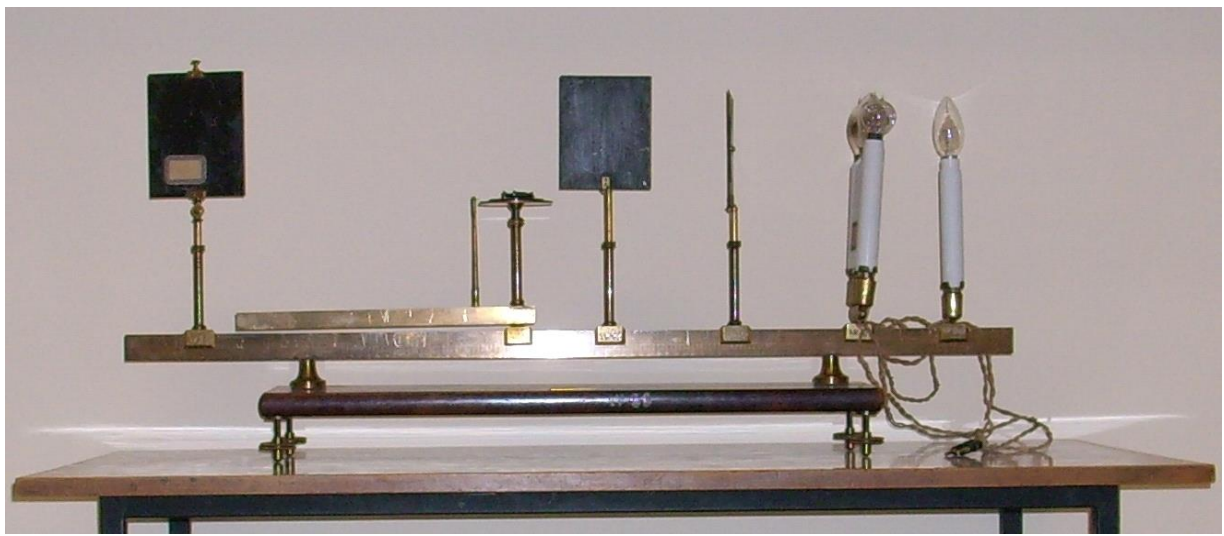

**Fig. 5.** Foto del banco ottico dopo gli interventi di restauro.

## 4. Discussione e conclusioni

Il banco ottico fu costruito da Filippo Caliri, molto probabilmente, prendendo a modello quello del Gabinetto di Fisica dell'Università di Palermo, acquistato nel 1850 dalle officine Ruhmkorff di Parigi (Inv. N. 184 del 1850). Considerato inoltre che la sorgente luminosa è costituita da una serie di lampade a incandescenza alimentate elettricamente, si può ragionevolmente pensare che lo strumento sia stato costruito tra la fine del XIX e l'inizio del XX secolo, subito dopo l'elettrificazione della città di Palermo. Purtroppo, non è stato possibile trovare riferimenti negli inventari del Liceo, in quanto molto materiale soprattutto cartaceo è andato distrutto durante i bombardamenti dell'ultimo conflitto mondiale.

Oggi il Liceo "Umberto I", consapevole del ruolo qualificante che gli strumenti scientifico-didattici rivestono nei processi di formazione scolastica e della loro valenza storica, ha intrapreso un rilevante percorso propedeutico all'istituzione di un museo scolastico tematico, al fine di garantire la salvaguardia delle collezioni scientifiche e favorire un processo di diffusione della conoscenza, riguardo la storia del Liceo e delle sue collezioni (Lo Scrudato, 2015). In tale prospettiva, la collaborazione con l'Università consente di creare per gli studenti nuove opportunità di apprendimento extra-scolastico e di crescita con valenza didattica per quanto riguarda lo studio delle leggi fisiche che stanno alla base del funzionamento di uno strumento scientifico e con valenza professionalizzante per quanto riguarda le tecniche e le metodologie di restauro, contribuendo in questo modo attivamente alla valorizzazione e tutela della collezione del Liceo.

In conclusione, nell'articolo abbiamo discusso del recupero di un raro banco ottico del Melloni, appartenente alla collezione di strumenti scientifici del Liceo Classico Statale "Umberto I" di Palermo. Gli interventi effettuati hanno permesso di individuare la costruzione di questo banco ottico presso l'Officina Tecnica Piazzi, fondata da Filippo Caliri nel 1870 a Palermo, fornendo un elemento di rarità che valorizza non solo lo strumento restaurato ma l'intera collezione di cui esso fa parte. Gli studenti, che hanno frequentato lo Stage Formativo, organizzato dal Dipartimento di Fisica e Chimica, hanno migliorato il loro apprendimento in riferimento agli argomenti scientifici connessi con l'uso del banco ottico per la dimostrazione delle leggi fisiche sul calore raggiante (radiazione infrarossa) e la sua analogia con la luce.

**Bibliografia e sitografia**

CALIRI F., MACALUSO D., 1871. Misure barometriche delle altezze sul livello del mare di talune montagne che circondano Palermo. Giornale di scienze Naturali ed Economiche Vol. VII, Parte I, 138.

GANOT A., 1863. Trattato elementare di fisica sperimentale ed applicata e di meteorologia. Pagnoni Editore, XVII Edizione, Milano, p. 314.

Gazzetta Ufficiale del Regno d'Italia N. 202 del 24 Agosto 1914

LO SCRUDATO V., 2015. Storia del Liceo "Umberto I" di Palermo attraverso la dotazione tecnico scientifica. In: Agliolo Gallitto A. (ed.), Atti del Convegno, Gli strumenti scientifici delle collezioni storiche nell'area palermitana, Palermo 23 - 24 ottobre 2014. Quaderni di Ricerca in Didattica (Science), suppl. n. 7: 11-18.

MAROTTI R., 2004. Introduzione al restauro della strumentazione di interesse storico scientifico. Il Prato, Padova.

MINIATI M., 1988. Il restauro degli strumenti scientifici. Alinea ed., Firenze.

SESSA E., 2008. Arte e architettura in Sicilia fra «belle époque» e «anni ruggenti». In: Quartarone C., Sessa E., Mauro E., Arte e architettura liberty in Sicilia, 131-170.